\documentclass{article}
\usepackage{shadethm}
\newshadetheorem{definition}{Definition}
\newshadetheorem{observation}{Observation}
\usepackage{marvosym}
\newcommand{\bhvr}[1]{\hbox{${\beta_{\scriptsize #1}}$}}
\newcommand{\bhvrtwo}[2]{\hbox{${\beta^{\scriptsize #2}_{\scriptsize #1}}$}}
\def\M{\textnormal{\textbf{M}}}
\def\A{\textnormal{\textbf{A}}}
\def\P{\textnormal{\textbf{P}}}
\def\E{\textnormal{\textbf{E}}}
\def\K{\textnormal{\textbf{K}}}
\def\OO{\textnormal{\textbf{O}}}
\usepackage{pbox}
\usepackage{latexsym}
\usepackage{amsmath}
\usepackage{amssymb}
\usepackage{graphicx}
\usepackage[pdftex,
            pagebackref=true,
            colorlinks=true,
            linkcolor=blue
           ]{hyperref}
\usepackage{float}
\usepackage{alltt}
\usepackage{authblk}

\begin{document}

\title{On Resilient Behaviors\\in Computational Systems and Environments}
\author{Vincenzo De Florio}
\affil{MOSAIC research group\\
       iMinds research institute \& University of Antwerp\\
       Middelheimlaan 1, 2020 Antwerp, Belgium\\
       \texttt{vincenzo.deflorio\MVAt{}gmail.com}}
\date{}
\maketitle
\begin{abstract}
The present article introduces a reference framework for discussing resilience of computational systems.
Rather than a property that may or may not be exhibited by a system,
resilience is interpreted here as the emerging result of a dynamic process.
Said process represents the dynamic interplay between
the behaviors exercised by a system and those of the environment it is set to operate in.
As a result of this interpretation, coherent definitions of several aspects of resilience can be
derived and proposed, including elasticity, change tolerance, and antifragility.
Definitions are also provided for measures of the risk of unresilience as
well as for the optimal match of a given resilient
design with respect to the current environmental conditions.
Finally, a resilience strategy based on our model is exemplified through a
simple scenario.
%
\end{abstract}

\section{Introduction}

Resilience is one of those ``general systems attributes'' that appear to play a central role
in several disciplines. Examples include
              ecology, business, psychology, industrial safety~\cite{laprie:dsn08},
microeconomics, computer networks, security, management science, cybernetics,
              control theory, as well as crisis and disaster management and
              recovery~\cite{DF14d,5361311,Salles19102011,Meyer:09,DFSB14}.
Although common traits are retained, in each discipline resilience takes peculiar 
do\-main-specific meanings~\cite{DF14d}. To exacerbate the problem, even in the context of
the same discipline often no consensus has been so far reached as to what are the
peculiar aspects of resilience and what makes it different from, e.g., elasticity,
dependability, or antifragility.

The present article contributes towards a solution to this problem in several ways.
First, in Sect.~\ref{s:mod}, we introduce resilience, compare
various of its domain-specific definitions, and derive a number of ancillary
concepts and working assumptions. This allows
resilience to be interpreted
in Sect.~\ref{s:bea}
      as the property emerging from the interaction
      of the behaviors exercised by a system and the environment it is set
      to operate in.
The outcome of said interaction depends on both intrinsic and extrinsic
factors: the ``traits'' of the system together with its endowment---the system's
peculiar characteristics
as well as its current state and requirements. At stake is the \emph{identity\/} of the
system, which we identify here with compliance to system specifications, including
functional and non-functional system requirements.

The resilience interaction
is modeled by considering the behaviors produced by the system and its environment.
This provides us with a unifying framework within which it is possible to express coherent definitions
of concepts such as elasticity, entelechism (change tolerance), and antifragility.
      Both system and environment are further modeled in terms of the
      ``resilience organs'' managing the five major services
      ancillary to resilience~\cite{DF13a,DF13b}: perception, apperception, planning,
      executive, and knowledge management organs---cor\-res\-pon\-ding to the
      five modules of autonomic systems~\cite{KeCh:2003}.
%
After this, in Sect.~\ref{s:app}, we introduce measures of the optimality of a given resilient
design with respect to the current environmental conditions: system supply and
system-environment fit.
One such strategy is detailed and exemplified in Sect.~\ref{s:sce} through an ambient intelligence
scenario.
%
%
%
%
Finally, major results are recalled and final conclusions
are drawn in Sect.~\ref{s:end}.

\section{Basic Concepts}\label{s:mod}


The term ``resilience'' comes from Latin resil{\=\i}re, ``to spring back, start back, rebound, recoil, retreat'',
and is often intended and defined as the ability to cope with or recover from change.
As mentioned in Sect.~1, this general definition has been specialized
in different domains, in each of which it has taken domain-specific traits:
\begin{itemize}
\item In ecology, resilience often refers to an ecosystem's ability to respond to and recover
      from an ecological threat~\cite{Holling73}. ``Recovering'' means here the ability
      to return to a steady state characterizing the ecosystem before the manifestation of the
      threat. The mentioned ``steady state''
      represents the peculiar characteristics of the resilient ecosystem---its \emph{identity}.
\item In complex socio-ecological systems, resilience is the
      ability to absorb stress and maintain function in the face of climate change~\cite{Folke06}.
      ``Absorbing stress'' clearly corresponds to the ``recovering'' ability found in ecology, though
      in this context the identity of the system lies in its function rather than in its state.
      An additional and peculiar aspect of resilient systems in this domain is given by their ability to improve
      systematically their sustainability.
\item In organization science, (organizational) resilience is ``the capacity
      to anticipate disruptions, adapt to events, and create lasting value''~\cite{ResCorp}.
      Here the accents are on proactiveness and adaptation rather than on ``springing back''
      to a past state or function. Intuitively, one may deduct that the former class
      of behaviors is more advanced than the latter one. One may also observe how
      in this case the definition brings to the foreground a fundamental component of system identity,
      namely the ability to create value.
\item In social science and human ecology, resilience is ``the ability of groups or communities to cope with
      external stresses and disturbances as a result of
      social, political and environmental change''~\cite{Adger00}. Here the recovery strategy of resilience
      is not made explicit. System identity implicitly refers to the identity of groups and communities and
      ranges from socio-cultural aspects up to the ability to survive.
\item In psychology, ``resilience is the process of adapting well in the face of adversity,
      trauma, tragedy, threats or significant sources of stress [...] It means `bouncing back' from
      difficult experiences''~\cite{APAresil}. An important aspect here is the identification of resilience
      as a \emph{process}: ``Resilience should be considered a process, rather than a trait to be had''~\cite{Rutter08};
      see also~\cite{Masten94,gilligan2001promoting}.
      ``System'' identity is in the case of psychology the collection of beliefs about oneself.
\item In material science, resilience is a material's property to ``stay unaffected by an
      applied stress'' up to some threshold, called ``point of yielding'', and to ``return to its
      original shape upon unloading''~\cite{Roylance2001}. Beyond the mentioned threshold,
      deformation is irreversible: ``some residual strain will persist even after unloading'' (ibid.)
      Here resilience is a property rather than a process, the key difference being the
      type of behavior exercised by the ``systems''.
      System identity is in this case represented by the shape or the characteristics of the material.
\item In civil engineering, resilience is a construction's ability to absorb or avoid damage in
      the face of a natural or man-induced abnormal condition~\cite{Jennings13}, such as flooding,
      hurricanes, or firepower. The considerations done in the case of material science apply
      also to this case.
\item Finally, in computer science, resilience has been defined, e.g., as 
      the ability to sustain dependability when facing changes~\cite{laprie:dsn08}.
      This translates into the ability to avoid failure and at the same time the ability to sustain the delivery
      of trustworthy services. System identity is in this case
      as in the following definition.
\end{itemize}

\begin{definition}[System identity]\label{d:id}
In the framework of artificial, computer-based systems,
system identity is defined as a system's compliance to its
system specifications and in particular to its functional and
non-functional quality of service and quality of experience requirements. 
\end{definition}


As already remarked, with the change of the reference domain
the above notions of resilience 
are applied to a spectrum of entities ranging from simple, passive-behaviored, individual objects
to complex, teleological, collective adaptive systems.
By making use of the behavioral approach introduced by Wiener et al. in~\cite{RWB43},
and briefly recalled in Sect.~\ref{s:bea:rwb},
in what follows three major classes in this spectrum are identified.

\subsection{Elastic objects and systems}\label{s:mod:res:ela}

Resilience shall be referred to as elasticity when the system under consideration is
only capable of simple types of behaviors:
passive behavior and active, purposeful, non-teleological behaviors. In the former case the system
shall be referred to as an object.

The considerations made above with reference to, e.g., material science,
apply also in this case. In particular,
for both objects and servo-mechanisms resilience (elasticity) is represented
as an \emph{intrinsic\/} property: a trait.

Elastic systems able to exercise active behaviors are what Boulding refers to
in~\cite{Bou56} as ``servo-mechanisms''---systems whose action is predefined and
is not modified by the interaction with other systems. In fact servo-mechanisms 
do not ``interact with other systems outside of themselves''~\cite{Hey98}---namely,
they are not open systems.

Elastic systems and objects
operate under the assumption that their environments
are not going to exercise stress beyond their point of yielding.
Quoting N.~N. Taleb, they are systems that ``do not care too much'' about their

Another way to characterize elastic objects and systems is by observing that they have
a predefined and static point of yielding. This introduces two syndromes, which we call
``elastic undershooting'' and ``elastic overshooting.''

\subsubsection{Elastic undershooting}
Elastic objects or systems are characterized at design time by
a static yielding point beyond which they permanently lose their identity---for instance, they deformate;
or break down; or fail; or become untrustworthy.
The yielding point is therefore a \emph{resilience threshold}. Whatever the
characteristics of an elastic object or system, there is always a non-zero probability that
the yielding point will be overcome.
\begin{definition}[Elastic undershooting]
When at time $t$ an elastic object or system with yielding point $Y$ is insufficiently
resilient with respect to
an experienced condition for which a yielding point $y(t)$ would be required,
we shall call elastic undershooting (or simply undershooting, when this may be done
without introducing ambiguity)
the dynamic quantity $y(t)-Y$.
\end{definition}

Undershooting is in fact as in a well-known fairy tale~\cite{Jacobs1890}:
one may make their house of straw, of wood, or even of bricks; although more and more robust, each house
will ``just'' shift the yielding point farther away; but that is all: there is no guarantee that, sooner or
later, something stronger will show up and blow that house down, whatever the material it is made of.
Development and operating costs, on the other hand, will grow up proportionally to the
chosen yielding point---which brings us to the second syndrome.

\subsubsection{Elastic overshooting}\label{ss:elasticOverShooting}
The choice of the yielding point represents the
ability to cope with a worst-case scenario regardless of how frequently said condition
will actually manifest itself. Unless the environmental conditions are deterministic
and immutable, there will always be a non-zero probability that the yielding point is
more pessimistic than what the experienced condition would require. 
In other words,
an elastic system is prepared for the worst; but also it costs and expends resources as if the worst
was actually there \emph{all the time}.
\begin{definition}[Elastic overshooting]
When at time $t$ an elastic object or system with yielding point $Y$ is resilient with respect to
an experienced condition for which a yielding point $y(t)$ would suffice,
we shall call elastic overshooting
(or simply overshooting, when this may be done
without introducing ambiguity)
the quantity $Y-y(t)$.
\end{definition}

Overshooting reminds of the condition of the shell-snail that,
``feeling always in danger of birds, lives constantly under its shield''~\cite[p.~147]{Bayley1836}---thus
carrying its weight at all times regardless of the actual presence or absence of birds.

\subsubsection{Observations}
Elastic undershooting and overshooting may be better understood when considering a well-known result by Shannon~\cite{Shannon}:
given an unreliable communication channel, it is always possible to transfer information reliably through it
provided that a sufficient amount of information redundancy is foreseen. By means of the above introduced terminology,
Shannon's result may be formulated as follows: for any communication channel whose observed unreliability is $y(t)$
throughout a given time interval $T$, it is
possible to define an elastic communication protocol with a yielding point
\[ Y>y(t) \,\,\forall t\in T.\]
Ideally the choice of $Y$ should be such that
$Y$ represents the supremum of all the unreliability samples $y(t)$ observed during $T$.
In this ideal case, no undershooting is experienced, although the system exhibits a cumulative
overshooting equal to 
\[\int_T Y-y(t)dt.\]
In more concrete situations in which unreliability drifting is unbound, undershooting would occur
each time the value chosen for $Y$ would be less than the observed unreliability of the channel.


\subsection{System entelechies}\label{s:mod:res:ent}
In Sect.~\ref{s:mod} we have concisely reviewed a number of definitions of resilience
emerged in the framework of diverse disciplines and domains. Several of such definitions
explicitly require a resilient system to enact complex forms of behaviors: adapt reactively (see,
e.g., in ecology and
psychology) and adapt proactively (see, e.g., in organizational science). Such behaviors correspond respectively
to simple teleological behaviors and extrapolative teleological behaviors (as defined in~\cite{RWB43}
and recalled in Sect.~\ref{s:bea:rwb}):
behaviors that are driven respectively by the current state and by the hypothesized future state
of an intended objective.
Obviously in this case resilience cannot be regarded as a trait or attribute; rather, it is the emerging result
of a process. Resilient systems are \emph{in motion\/} to actively pursue the \emph{persistence of their
system identity}. The two just mentioned aspects correspond to the translation that Joe Sachs
provides of the Aristotelian concept of entelechy\footnote{%
	Quoting Sachs,
		``Entelecheia means continuing in a state of completeness,
		  or being at an end which is of such a nature
		  that it is only possible to be there by means of the
		  continual expenditure of the effort
		  required to stay there.''~\cite{SachsM}}:
``being-at-work-staying-itself''~\cite{Sachs,SachsM}.
An entelechy is a system that is able to persist and sustain its
\emph{completeness\/} through a resilient process. ``Completeness'' here is to be intended as
the characteristics that make of a system what it is: its ``definition''---or,
in other words, its system identity. Because of this we shall refer to the systems in this
resilience class as to \emph{entelechies}.

The very nature of entelechies requires them to be able to ``interact with other systems
outside of themselves'', namely to be open systems~\cite{Hey98}. Such systems do not ``want tranquility''
nor expect their environments to be stable or stay the same. On the contrary, they assume
conditions \emph{will\/} vary, and adjust their function to the observed conditions
or to speculated future conditions of their environments.

Another way to distinguish entelechies from elastic objects and systems
is by observing that entelechies are characterized by dynamic and adaptive points of yielding.
Undershootings and overshootings are still possible, though with a slightly different
formulation:

\begin{definition}[Entelechial undershooting]
When, at time $t$, an entelechy
with yielding point $Y(t)$ is insufficiently resilient with respect to an experienced
condition for which a yielding point $y(t)$ would be required, we shall call entelechial undershooting
(or simply overshooting, when this may be done
without introducing ambiguity)
the dynamic quantity $y(t) - Y(t)$.
\end{definition}

\begin{definition}[Entelechial overshooting]
When, at time $t$, an entelechy with yielding point $Y(t)$ is resilient with respect to
an experienced condition for which a yielding point $y(t)$ would suffice,
we shall call entelechial overshooting
(or simply overshooting, when this may be done
without introducing ambiguity)
the quantity $Y(t)-y(t)$.
\end{definition}

An exemplary entelechy is given by an adaptive
communication protocol for the reliable communication over
an unreliable channel characterized by $y(t)$ unreliability.
Such protocol would continuously ``be at work'' so as to estimate past and current
values of $y(t)$ and extrapolate with them a future state $y(t')$.
Once this speculated future value is known, the protocol would ``stay itself'' by
choosing a yielding point $Y(t')$
as close as possible but still greater than $y(t')$.

More formally, the choice for $Y(t')$ would be such that
\begin{equation}
0< Y(t')-\Pi(y(t'))<\varepsilon,\label{e:psilon}
\end{equation}
where $\Pi(y(t'))$ represents a prediction of $y(t')$ and $\varepsilon>0$ expresses a safety margin
to cover for inaccuracies in the prediction.

\subsection{Antifragile systems}\label{s:mod:res:ant}
In the light of the discussion in Sect.~\ref{s:mod:res:ela} and Sect.~\ref{s:mod:res:ent}
one may observe that
most of the reported definitions of resilience correspond to either elastic objects~/~systems or to entelechies.
An exception may be found in the class of
complex socio-ecological systems. There we have systems that 
``are at work to stay themselves'' (thus, they are entelechies), though
are endowed with an additional feature:
the ability ``to improve systematically their sustainability''.
Wiener et al. did not explicitly consider behaviors including that ability~\cite{RWB43}.
The most closely related of their behavioral classes---one could say its genus proximum~\cite{Burek04}---is
given by teleological behaviors.

As discussed in more detail in Sect.~\ref{s:bea:rwb}, teleological
systems are those characterized by a feed-back loop: their
behavior
\begin{quote}
``is controlled by the margin of error at which
the [system] stands at a given time with reference to a relatively specific goal''~\cite{RWB43}.
\end{quote}
Due to its purely behavioral nature, the approach followed in the cited work
does not cover organizational, architectural, and structural aspects. Because of this,
no account is given on the modifications that a teleologically behaviored
system would apply to itself in order to achieve its goal.

At least the following four cases may occur:
\begin{enumerate}
\item The feed-back loop is purely exogenous: the system action is simply steered
towards the goal (in its current or hypothesized future position.)
\item The feed-back loop is both exogenous and endogenous. Internal changes
only concerns the ``knobs'', namely the parameters of the system.
This case corresponds to parametric adaptation.
\item The feed-back loop is both exogenous and endogenous; the internal changes
adapt the structure of the system. This corresponds to system reconfigurations
(namely structural adaptation). Adaptations are \emph{phenotypical\/} and do not
affect the identity of the system.
Furthermore, the experience leaves no trace on the identity of the system.
\item The feed-back loop is both exogenous and endogenous; the internal changes adapt any
of the following aspects: the function;
the structure; the architecture; and the organization of the system. Changes
are \emph{genotypical}: they are persisted and modify permanently the nature of the system.
\end{enumerate}

It is important to remark that,
while in cases 1--3 the system is ``at work to stay itself''~\cite{Sachs,SachsM},
in case 4 the system is ``at work to get better''.
At least in the case of complex socio-ecological systems, teleological behaviors
belong to this fourth category:
through their experience, those systems elaborate
a feed-back that is also endogenous and affects the genotypical
ability ``to improve systematically their sustainability''. The feed-back thus
affects the identity of the system. Rather than adapting,
the system \emph{evolves}\footnote{A distinctive characteristic of both behavioral elasticity and entelechism
	is their goal being the persistence of system identity. Now, a strict interpretation
	of this property translates in prohibiting
	any \emph{evolution\/} of the system identity. When looking at natural systems we observe that
	two concurrent ``forces'' are used in biological evolution:
	genealogical persistence of identity and occasional identity disruption through mutation
	and Darwinian selection~\cite{HT:TE14a,Hol95}.
	Those forces are \emph{both\/} used to lead species and ecosystems to build ever greater amounts of
	complexity and thus to the ``evolution of evolvability'',
	i.e., ``the ability
	of random variations to sometimes produce improvement.''~\cite{WaAl1996}}.

In the case of complex socio-ecological systems, said evolution leads to
an improvement in sustainability: those systems 
``enhance the level of congruence or fit between themselves
and their surroundings''~\cite{stokols2013enhancing}.
This matches the concept
introduced by N. N. Taleb in~\cite{Taleb12}: antifragile systems. Quoting from the
cited reference,
``Antifragility is beyond resilience or robustness. The resilient resists shocks
and stays the same; the antifragile gets better.''

In what follows we distinguish explicitly this class of teleological behaviors
and systems by referring to them as to \emph{antifragile systems},
which we define as follows:

\begin{definition}[Antifragile system]
We shall call a system ``antifragile'' if it is able to exercise teleological behaviors
that evolve the system and its identity in such a way as to systematically
improve the fit with their environment.
\end{definition}

By considering the just enunciated definition we can observe that 

\begin{itemize}
\item Antifragile systems are not necessarily more resilient
that entelechies or elastic objects and systems. As it is the
case for those entities, also antifragile
systems are characterized by a yielding point---a resilience threshold beyond
which they would fail; break down; or become untrustworthy.
\item Antifragile systems mutate their system identity. By referring to Def.~\ref{d:id},
this means that the behaviors of antifragile computer-based systems may drift outside of what prescribed
in their specifications. Scenarios such as those that Stephen Hawking~\cite{Hawking14}
and many others~\cite{OpenLetter} are warning of
become more concrete when considering
this particular characteristic of antifragile systems.
\item Antifragile systems must possess some form of awareness of their current and past
system-environment fit; in particular, they must
be able to create and maintain a model of the risk of losing
one or more aspects of their system identity.
\end{itemize}

Going back to the communication protocol presented in Sect.~\ref{s:mod:res:ent},
an exemplary antifragile system would be a 
protocol that, in addition to being able to estimate
quantity $y(t')$, also learns how to mutate its own algorithm so as to profit
from the characteristics of the environment. As an example,
instead of sending, say, $Y$ redundant copies for each of the
packets of its messages, the protocol could
realize that a better strategy (with respect to the current behavior
of the channel) would be that of interleaving the transmission
of packets of different messages. This would result in a more efficient strategy such
that a higher yielding point would be reached with a consumption of resources
lower than in the original algorithm.

At the same time, it is important to observe how the introduced interleaving would affect
several peculiar characteristics of the protocol---for instance, it would
introduce jitter (viz. a drifting in the periodicity of the messages).
Repercussions on the validity of the specifications become then possible.
For instance,
if the protocol were intended for a teleconferencing service, the
introduced jitter would affect the quality of experience of the users
of that service. Embedding the same protocol in a service insensible
to periodicity drifting (such as a
file transferring service) would not translate in a loss of system identity.



\subsection{A few observations}
As a summary of the discussion in this section, we can derive here a number of observations:

\subsubsection{Resilience is a relative figure}\label{ss:relative}
As observed in~\cite{DF14b}, resilience is a dynamic property
whose emergence is influenced by at least the following two factors:
\begin{enumerate}
\item The intrinsic characteristics of the system: in particular, whether the system
is elastic, entelechial, or antifragile.
\item The extrinsic ``level of congruence or fit between [the system]
and [its] surroundings''~\cite{stokols2013enhancing}.
\end{enumerate}
The first factor is absolute in nature and tells how ``evolved'' the class of the system is.
The second factor is a relative one, and tells how the system's behavior is able to match the
conditions currently expressed by the environment.
This second factor makes of resilience a relative figure.
Whatever a system's structure, organization, architecture, capabilities, and resources, that system
is resilient only so long as its implementation matches the conditions currently exercised by
the environment\footnote{Possibly the first Scholar to have distinguished
	intrinsic and extrinsic factors towards
	the emergence of resilience was G.~W. von Leibniz~\cite{DF14c}.}.

\subsubsection{Resilience is the product of an interplay between a system and its environment}
As a corollary to what mentioned in Sect.~\ref{ss:relative}, we observe that
resilience is not \emph{a property}, but rather the product of a \emph{a process}.
Such process corresponds to the dynamic interplay between two 
entities: A system and its environment.

\subsubsection{The environment is a system}
``Environment'' is interpreted here and in what follows simply as
      another system; in particular,
      as a \emph{collective\/} system (in other words, a ``system-of-systems'')
      taking different shapes,
      including for instance any combination of cyber-physical things;
      biological entities such as human beings; servo-mechanisms;
      and intelligent ambients able to exercise complex teleological behaviors.

\subsubsection{Resilience is an interplay of behaviors}
Resilience is one of the possible outcomes of an interplay of behaviors. 
      If and only if the interplay between the system behaviors and the environmental behaviors
      is one that preserves the
      system identity then the system will be called resilient.
As discussed in Sect.~\ref{s:bea:rwb}, behaviors may range
from the random behaviors typical of
electro-magnetic sources up to the ``intelligent'', cybernetic behaviors characterizing, e.g.,
human beings and complex ambient environments.

\subsection{Preliminary conclusions}
A major conclusion here---and a starting point for the treatise in next section---is given by the intuitive notion
that evaluating resilience must be done not merely considering a system's intrinsic characteristics; rather,
it should be done by expressing in some convenient form the dynamic fit between the system and its environment.
This may be obtained, e.g., by comparing the resilience class of the system with the dynamically mutating
resilience class of the environment. Another, more detailed method could be by comparing the \emph{behaviors\/}
of system and environment. Yet another approach could be to apply the behavioral comparison method
to specific \emph{organs\/} of the system and its environment---for instance those organs that are
likely to play a significant role in the emergence of resilience or its opposite.

In what follows we focus on the last mentioned approach.

\section{Resilient Behaviors, Organs, and Methods}\label{s:bea}
In order to proceed with the present treatise we first recall in Sect.~\ref{s:bea:rwb}
the major classes of behaviors according to the classic
discussion by Rosenblueth, Wiener, and Bigelow~\cite{RWB43}.
After this, in Sect.~\ref{s:bea:pap},
five major services that play a key role towards the emergence of resilience
are identified.
Finally in Sect.~\ref{s:bea:def} we use the concepts introduced so far to reformulate definitions
for elasticity, entelechism, and antifragility.

\subsection{Behavioral classification}\label{s:bea:rwb}
As already mentioned, Rosenblueth, Wiener, and Bi\-ge\-low introduced in~\cite{RWB43}
the concept of
the ``behavioristic study of natural events'', namely ``the examination of the
output of the object and of the relations of this output to the
input''\footnote{If not otherwise specified the quotes in
                 the present section are from~\cite{RWB43}.}.
The term ``object''
in the cited paper corresponds to that of ``system''.
In that renowned text the authors purposely 
``omit the specific structure and the intrinsic organization'' of the systems under scrutiny
and classify them exclusively on the basis of the quality of the ``change produced
in the surroundings by the object'', namely the system's behavior.
The authors identify in particular four major classes of behaviors\footnote{For the sake
	of brevity passive behavior shall not be discussed here.}:
\begin{description}
	\item[\bhvr{\hbox{ran}}\,:] Random behavior. This is an active form of behavior that does not appear to serve
		a specific purpose or reach a specific state. A source of electro-magnetic interference 
		exercises random behavior.
	\item[\bhvr{\hbox{pur}}\,:]
		Purposeful, non-teleological behavior. This is behavior that serves a purpose and is directed towards
		a specific goal. In purposeful behavior a
		``final condition toward which the movement [of the object] strives'' can be identified.
                Servo-mechanisms provide an example of purposeful behavior.
	\item[\bhvr{\hbox{rea}}\,:]
		Reactive, teleological, non-evolutive behavior. This is behavior that
		``involve[s] a continuous feed-back from the goal that modifies and
		guides the behaving object''. 
		Examples of this behavior include phototropism, namely the tendency that can
                be observed, e.g., in certain plants,
		to grow towards the light, and gravitropism, viz. the tendency of plant roots to grow downwards.
		As already mentioned (see Sect.~\ref{s:mod:res:ent}),
		reactive behaviors require the system to be open~\cite{Hey98} (i.e., able to
		continuously perceive, communicate, and interact with external systems and the
		environment) and to embody some form of feedback loop.
		Class \bhvr{\hbox{rea}} is non-evolutive, meaning that the experienced change does
		not influence the identity of the system (cf. Sect.~\ref{s:mod:res:ant}).
	\item[\bhvr{\hbox{pro}}\,:]
		Proactive, teleological, non-evolutive behavior.
		This is behavior directed towards the extrapolated future state of the goal.
		The authors in~\cite{RWB43} classify proactive behavior according to its ``order'', namely
		the amount of context variables taken into account in the course of the extrapolation.
		As class \bhvr{\hbox{rea}}, so class \bhvr{\hbox{pro}} is non-evolutive.
\end{description}

By considering the arguments in Sect.~\ref{s:mod:res:ant} a fifth class can be added:
\begin{description}
	\item[\bhvr{\hbox{ant}}\,:] teleological evolutive behaviors.
	This is the behavior emerging from
	antifragile systems (see Sect.~\ref{s:mod:res:ant} for more detailed
	on this class of systems).
\end{description}

Each of the above five classes may see their systems operate in isolation
or through some form of social interaction. In order to differentiate
these two cases we add the following attribute:

\begin{description}
	\item[$\sigma(b)$\,:]
		True when $b$ is a social behaviors.
		This attribute identifies behaviors based on the social
		interaction with other systems deployed in the same environment.
		Examples of such behaviors include, among others, mutualistic,
		commensalistic, parasitic, co-evolutive, 
		and co-opetitive behaviors~\cite{BN98,AF83}. For more information
		the reader is referred to K. Boulding's discussion in
		his classic paper~\cite{Bou56} and, for a concise survey
		of social behaviors, to~\cite{DF14d}.
\end{description}

The resilience classes introduced in Sect.~\ref{s:mod} can now be characterized in terms of the above
behavioral classes: elastic systems correspond to \bhvr{\hbox{pur}}; non-evolving entelechies
exercise either \bhvr{\hbox{rea}} or \bhvr{\hbox{pro}} behaviors; while, as already mentioned,
\bhvr{\hbox{ant}} pertains to antifragile systems.

We shall define $\pi$ as a projection map returning, for each of the above behavior classes, 
an integer in $\{1,\ldots,5\}$ ($\pi(\bhvr{\hbox{ran}})=1$, \ldots, $\pi(\bhvr{\hbox{ant}})=5$).
Aim of $\pi$ is twofold: it associates an integer ``identifier'' to each behavioral class
and it introduces an ``order'' among classes. Intuitively, the higher is the order of class,
the more complex is the behavior.

In what follows it is assumed that behaviors manifest themselves by changing
the state of measurable properties. As an example, behavior may translate into
a variation in the electromagnetic spectrum perceived as a change in luminosity
or color.
Context figures is the term that shall be used
in what follows to refer to those measurable properties.

For any behavior \bhvr{x}{} dependent on a set of context figures $F$, notation \bhvrtwo{x}{F}{} will be used to denote that \bhvr{x}{} is exercised
by considering the context figures in $F$. Thus if, for instance, $F=(\hbox{speed},\hbox{luminosity})$, then
\bhvrtwo{\hbox{rea}}{F}{} refers to a reactive behavior that responds to changes in speed and light.

For any behavior \bhvr{x}{} and any integer $n>0$, notation \bhvrtwo{x}{n}{} will be used to denote that \bhvr{x}{} is exercised
by considering $n$ context figures, without specifying which ones.

As an example,
behavior \bhvrtwo{\hbox{pro}}{|F|}{}, with $F$ defined as above,
identifies an 
order\mbox{-}\nobreak\hspace{0pt}2
proactive behavior, while \bhvrtwo{\hbox{pro}}{F}{} says in addition that that
behavior considers both speed and luminosity 
to extrapolate the future position of the goal.

Now the concept of partial order among behaviors is introduced.

\begin{definition}[Partial order of behaviors]\label{d:partial.order}
Given any two behaviors $\beta_1$ and $\beta_2$, $\beta_1 \prec \beta_2$ if and only if
either of the following conditions holds:
  \begin{enumerate}

  \item $\left(\pi(\beta_1) \le \pi(\beta_2)\right) \wedge
	 \left(\exists (F, G): \beta_1=\bhvrtwo{1}{F} \wedge
		\beta_2=\bhvrtwo{2}{G} \wedge F \subsetneq G\right)$.

	In other words, $\beta_1 \prec \beta_2$ if (1) $\beta_1$ belongs to a behavioral class that is
at most equal to $\beta_2$'s (via function $\pi$) and (2)
	$\beta_2$ is based on a set of context figures that extends $\beta_1$'s.
		\label{d:partial.order:two}
  \item $\left(\pi(\beta_1) \le \pi(\beta_2)\right) \wedge
	 \left(\exists (F, G): \beta_1=\bhvrtwo{1}{|F|} \wedge
		\beta_2=\bhvrtwo{2}{|G|} \wedge F \subsetneq G\right)$.

	This case is equivalent to case~\ref{d:partial.order:two}, the only difference being
	in the notation o the behavior.
  \item $\left(\pi(\beta_1) = \pi(\beta_2)\right) \wedge
	 (\sigma(\beta_1) = \hbox{false}) \wedge (\sigma(\beta_2) = \hbox{true})$.

	In other words, $\beta_1 \prec \beta_2$ also when 
	both $\beta_1$ and $\beta_2$ belong to the same behavioral class,
	though $\beta_2$ is a social behavior while $\beta_1$ is not.
  \end{enumerate}
\end{definition}

For any two resilient systems $p_1$ and $p_2$, respectively characterized by behavioral classes
$\beta_1$ and $\beta_2$, if
$\beta_1 \prec \beta_2$ then
$p_1$ is said to exhibit ``systemically inferior''\label{p:sysinf} resilience with respect
to $p_2$.

It is important to observe that $\prec$ is about the intrinsic resilience
characteristics of the system (see Sect.~\ref{ss:relative}). Partial order $\prec$
does not tell which system is ``more resilient''; it highlights
that for instance system ``dog'' is able to exercise behaviors that are less complex than
those of system ``man''. This tells nothing about the 
extrinsic ``level of congruence or fit''~\cite{stokols2013enhancing} that
for instance
a ``man'' or a ``dog'' may exhibit in a given environment.
As exemplified, e.g., in~\cite{Eraclios13.11.09}, when a threat is announced
by ultrasonic noise, a ``dog'' able to perceive the threat and flee could result
more resilient than a ``man''. The use of sentinel species~\cite{dS99+} is in fact a social
behavior based on this fact. An application of this principle is given
in Sect.~\ref{s:sce}.

\subsection{Resilience organs}\label{s:bea:pap}
As done in~\cite{DF14b} it is conjectured here
that reasoning about a system's resilience is facilitated by
considering the behaviors of the system organs that are responsible for the
following abilities:

\begin{description}
\item[\textbf{M}:] the ability to perceive change;
\item[\textbf{A}:] the ability to ascertain the consequences of change;
\item[\textbf{P}:] the ability to plan a line of defense against threats deriving from change;
\item[\textbf{E}:] the ability to enact the defense plan being conceived in step \textbf{P};
\item[\textbf{K}:] and, finally, the ability to treasure up past experience and systematically improve,
          to some extent, abilities \M, \A, \P, and \E.
\end{description}

These abilities correspond to the components of the so-called MAPE-K loop of autonomic
computing~\cite{KeCh:2003}.
In the context of the present paper the system components responsible for those
abilities shall be referred to as ``resilience organs''.

The following notation shall be used to refer to organ \OO{} of system $s$: $s.$\OO{} (for \OO$\in \{\M,\dots,\K\}$).

\begin{definition}[Cybernetic class]
For any system $s$ the 5-tuple corresponding
to the behaviors associated to its resilience organs,
\[(s.\mathbf{M}, s.\mathbf{A}, s.\mathbf{P}, s.\mathbf{E}, s.\mathbf{K}),\]
shall be referred to as the cybernetic class of $s$.
\end{definition}

Two observations are important for the sake of our discussion.
\begin{observation}[Intrinsic resilience]
A system's cybernetic class puts to the foreground how \emph{intrinsically resilient\/} that
      system is (see again Sect.~\ref{ss:relative})
 and makes it easier to compare whether certain resilience organs (or the whole system)
      are (resp., is) systemically inferior to (those of) another system.\label{o:intr}
\end{observation}
As an example,
the adaptively redundant data structures described in~\cite{DeBl08a} have the following cybernetic class:
\[ {\mathcal C}_1 = (\bhvr{\hbox{pur}}, \bhvrtwo{\hbox{pro}}{1}, \bhvr{\hbox{pur}}, \bhvr{\hbox{pur}}, \emptyset), \]
while the adaptive $N$-version programming system introduced in~\cite{BDB11a,BDFB12} is
\[ {\mathcal C}_2 = (\bhvr{\hbox{pur}}, \bhvrtwo{\hbox{pro}}{2}, \bhvr{\hbox{pur}}, \bhvr{\hbox{pur}}, \bhvr{\hbox{pur}}). \]
By comparing the above 5-tuples $\mathcal{C}_1$ and $\mathcal{C}_2$ one may easily realize how the major strength
of those two systems lies in their analytic organs, both of which are capable of proactive behaviors
(\bhvr{\hbox{pro}})---though in a simpler fashion in $\mathcal{C}_1$. Another noteworthy difference is the presence
of a knowledge organ in $\mathcal{C}_2$, which indicates that the second system is able to accrue and make use
of the past experience in order to improve---to some extent---its
action\footnote{The presence of a knowledge organ does not mean that its system is
	antifragile. In the case at hand, for instance, the system does not mutate it
	system identity---it stays an $N$-version programming system.}.

Please note that
not all the behaviors introduced in Sect.~\ref{s:bea:rwb} may be applied to all of a system's
organs. For instance, it would make little sense to have a perception organ behave randomly
(unless one wants to model, e.g., the effect of certain hallucinogenic substances in
chemical warfare\footnote{See for instance the interview with
	Dr. James S. Ketchum, in the Sixties a leading
	psychiatrist at the Army Chemical Center at Edgewood Arsenal in Maryland,
	US: 	``With BZ [3-quinuclidinyl benzilate], the individual becomes delirious,
		and in that state is unable to distinguish fantasy from reality, and may see,
		for instance, strips of bacon along the edge
		of the floor.''~\cite{HallucinogenicWeapons}}.)

\begin{observation}[Extrinsic resilience]
A system's cybernetic class puts to the foreground also how \emph{extrinsically resilient\/} that
      system is if the above comparison is done between the cybernetic class of the system and
      \emph{the dynamically evolving cybernetic class of the environment\/}
	(see Sect.~\ref{ss:relative}).
      This comparison represents a system-environment fit, in turn indicating the property emerging
      from the interplay between the current state of the system and
	the current state of the environment---in
      other words, whether the system is likely to either preserve or lose its peculiar
      features because of the interaction with the environment. In the former case
      the system shall be called as ``resilient;'' in the second case, ``unresilient.''\label{o:extr}
\end{observation}



\subsection{Again on elasticity, entelechism, and antifragility}\label{s:bea:def}
The model and approach introduced thus far provide us with a conceptual framework
for alternative definitions of elasticity, entelechism, and antifragility---namely
the resilience classes introduced in Sect.~\ref{s:mod}.

\begin{definition}[Elasticity]
Given a system $s$ and its cybernetic class $\mathcal{C}_s$, with $s$ deployed in environment $e$,
$s$ shall be called ``elastic''
with respect to $e$
if $s$ is resilient (in the sense expressed in Observation~\ref{o:extr}) and
if the behaviors in $\mathcal{C}_s$ are purposeful (\bhvr{\hbox{pur}})
and defined, once and for all, at design or deployment time.
\end{definition}
Elasticity corresponds to simple, static behaviors that 
make use of a system's predefined internal characteristics and resources
so as to \emph{mask\/} the action of external forces.
Those characteristics and resources take the shape of various forms of redundancy
which is used to mask, rather than tolerate, change.

\begin{definition}[Entelechism]
Given a system $s$ and its cybernetic class $\mathcal{C}_s$, with $s$ deployed in environment $e$,
$s$ shall be referred to as
``entelechy''
with respect to $e$
if $s$ is resilient (in the sense expressed in Observation~\ref{o:extr}) and
if the behaviors corresponding to the \A{} and \P{} organs of $\mathcal{C}_s$ are either
of
type \bhvr{\hbox{rea}} or \bhvr{\hbox{pro}}.
Entelechism, or change tolerance, is defined as the property exhibited by an entelechy.
\end{definition}
As evident from their definition, entelechies are open, context-aware systems;
able to autonomically repair their state in the face of disrupting changes;
and able to guarantee their system identity. A knowledge organ may or may not
be present and, depending on that, the feed-back organs may or may not be stateful---meaning
that ``memory'' of the experienced changes is or is not retained.

\begin{definition}[Computational antifragility]
Given a system $s$ and its cybernetic class $\mathcal{C}_s$, with $s$ deployed in environment $e$,
we shall say that $s$ is computationally
antifragile with respect to $e$ when the following conditions are all met:
\begin{enumerate}
\item The awareness organ of $s$, \A{}, is open to the system-environment fit between $s$ and $e$.
      In particular, as suggested in Sect.~\ref{s:mod:res:ant},
this means that \A{} implements a model of the risk of losing
      one or more aspects of the system identity of $s$. One such model
is exemplified, e.g., by the distance-to-failure function introduced
      in~\textnormal{\cite{DB07a}} and discussed
	in~\textnormal{\cite{DF13a}}.\label{d:anti:sefOpen}
\item Throughout time intervals in which the behavior of $e$ is stable, the planning organ \P{} is
      able to monotonically improve extrinsic resilience (see Obs.~\ref{o:extr})
and thus optimize risk/performance trade-offs.
      \label{d:anti:sefImprExtR}
\item Organ \P{} evolves through machine learning or other machine-oriented
      experiential learning~\textnormal{\cite{Kolb84}}, leading $s$ to evolve towards ever greater 
      intrinsic (systemic) resilience (see Obs.~\ref{o:intr}).
\item Organ \K{} is stateful and persists lessons learned from the experience and its ``conceptualization.''
\end{enumerate}
\label{d:anti}
\end{definition}

Computationally antifragile systems are system-env\-ir\-onment fit-aware systems;
able to embody and persist systemic improvements suggested by the match of the
current system identity with the current environmental conditions.
The learning activity possibly implies a \mbox{4-stage} cycle similar to the one
suggested by Kolb in~\cite{Kolb84}, executed concurrently with the resilience
behaviors of the \M, \A, \P, and \E{} organs.

\section{Approach}\label{s:app}

As already mentioned, a methodological assumption in the present article is that the evolution
of an environment may be expressed as
a behavior. Said behavior may be of any of the types listed in Sect.~\ref{s:bea:rwb} 
and Sect.~\ref{s:bea:def} and
it may result in the dynamic variation of a number of ``firing context figures''. In fact those figures characterize
and, in a sense, set the boundaries of an \emph{ecoregion}, namely
``an area defined by its environmental conditions''~\cite{Dictionary.com2014}.
An environment may be the behavioral result of the action of, e.g., 
a human being (a ``user''); or the software agents managing an intelligent ambient; or for instance
it may be the result of purposeless (random) behavior---such as a source of
electro-magnetic interference.
As a consequence, an environment may for instance behave (or appear to behave)
randomly, or it may exhibit a recognizable trend; in the latter case the
variation of its context figures may be such that it allows for tracking or speculation
(extrapolation of future states).
Moreover, an environment may exhibit the same
behavior for a relatively long period of time or it may vary frequently
its character.

\begin{figure}
	\centerline{\includegraphics[width=0.5\textwidth]{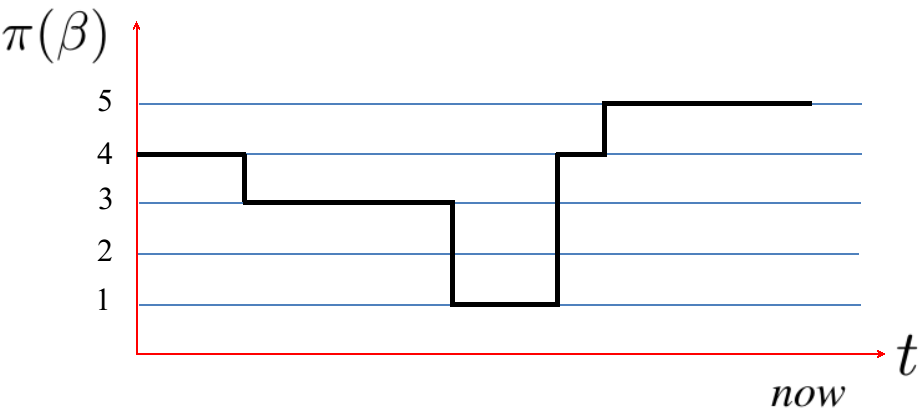}}
	\caption{Exemplification of turbulence, namely the dynamic evolution of environmental or 
	systemic behavior. Abscissas represent time,
	``\emph{now}'' being the current time. Ordinates are the behavior classes exercised
	by either the environment or the system.}\label{f:env}
\end{figure}

Given an environment (or a system), the dynamic evolution of the environmental
(resp. systemic) behavior shall be referred to in what follows as
``turbulence''.
Diagrams such as 
the one in Fig.~\ref{f:env} may be used
to represent the dynamic behavioral evolution of either environments or systems.

Whenever two behaviors $\beta_1$ and $\beta_2$ are such that $\beta_1 \prec \beta_2$, it is possible to
define some notion of distance between the two behaviors. One way to define
such ``behavioral metric function'' would be
by encoding the characteristics
of each behavior onto the bits of a binary word, with the three most significant bits encoding
the projection map of the behavior and the bits from the fourth onward encoding the
cardinality of the set of context figures or the order of the behavior.
When the behaviors belong to the same class although consider two different
context sets, say $F$ and $G$, then a simpler formulation of a distance would be
$\mathbf{abs}(|F|-|G|)$.
Let us call
$\mathbf{dist}$ one such metric function.


It is now possible to propose a definition of two indicators for the
extrinsic quality of resilience: The system supply relative to an environment
and the system-environment fit.

\begin{definition}[System supply]\label{d:supply}
	Let us consider a system $s$ deployed in an environment $e$, characterized respectively by
	behaviors $\beta^s(t)$ and $\beta^e(t)$.
	Let us assume that $\beta_1$ and $\beta_2$ are such that
	either $\beta_1 \prec \beta_2$ or $\beta_2 \prec \beta_1$.
	Given a behavioral metric function $\mathbf{dist}$ defined as above,
	the following value shall be called as supply of $s(t)$ with respect to $\beta^e(t)$:
	\begin{multline}
	\mathbf{supply}(s,e,t) = \nonumber \\
	= \begin{cases} 
		 \mathbf{dist}(\beta^s(t),\beta^e(t))  & \mbox{if } \beta^e(t)\prec \beta^s(t)\\
		-\mathbf{dist}(\beta^s(t),\beta^e(t))  & \mbox{if } \beta^s(t)\prec \beta^e(t)\\
		0 & \pbox{0.5\textwidth}{\mbox{if  $\beta^e(t)$ and $\beta^s(t)$}\\
					 \mbox{express the same behaviors.}}
	\end{cases}
	\end{multline}
\end{definition}

Supply can be positive (referred to as ``oversupply''), negative (``undersupply''),
or zero (``perfect supply'').

\begin{observation}
Oversupply and undersupply provide a quantitative formulation of the notions
of overshooting and undershooting given in Sect.~\ref{s:mod}.
\end{observation}

\begin{definition}[System-environment fit]\label{d:fit}
	Given the same conditions as in Definition~\ref{d:supply},
	the following function:
	\begin{multline}
		\mathbf{fit}(s,e,t) = \\ \nonumber \\
	= \begin{cases} 
		1 / (1 + \mathbf{supply}(s,e,t) ) & \mbox{if } \mathbf{supply}(s,e,t) \ge 0\\
		-\infty & \mbox{otherwise.}
	\end{cases}
	\end{multline}
	shall be referred to as ``system-environment fit of $s$ and $e$ at time $t$.''
\end{definition}

The above definition expresses system-environment fit as a function returning 1 in the case
of best fit; slowly scaling down with oversupply; and returning $-\infty$ in case of undersupply.
The reason for the infinite penalty in case of undersupply is due to the fact that
it signifies an undershooting or, in other words, a loss of system identity.

The just enunciated formulation is of course not the only possible one:
an alternative one could be, for instance,
by having $\mathbf{supply}^2$ instead of $\mathbf{supply}$ in the denominator of
$\mathbf{fit}$ in Def.~\ref{d:fit}. Another formulation could extend optimal fit to
a limited region of oversupply
as a safety margin to cover for inaccuracies in
the prediction of the behavior of the environment.
This is similar to the role of $\varepsilon$ in~\eqref{e:psilon}.

\begin{figure*}
	\centerline{\includegraphics[width=0.6\textwidth]{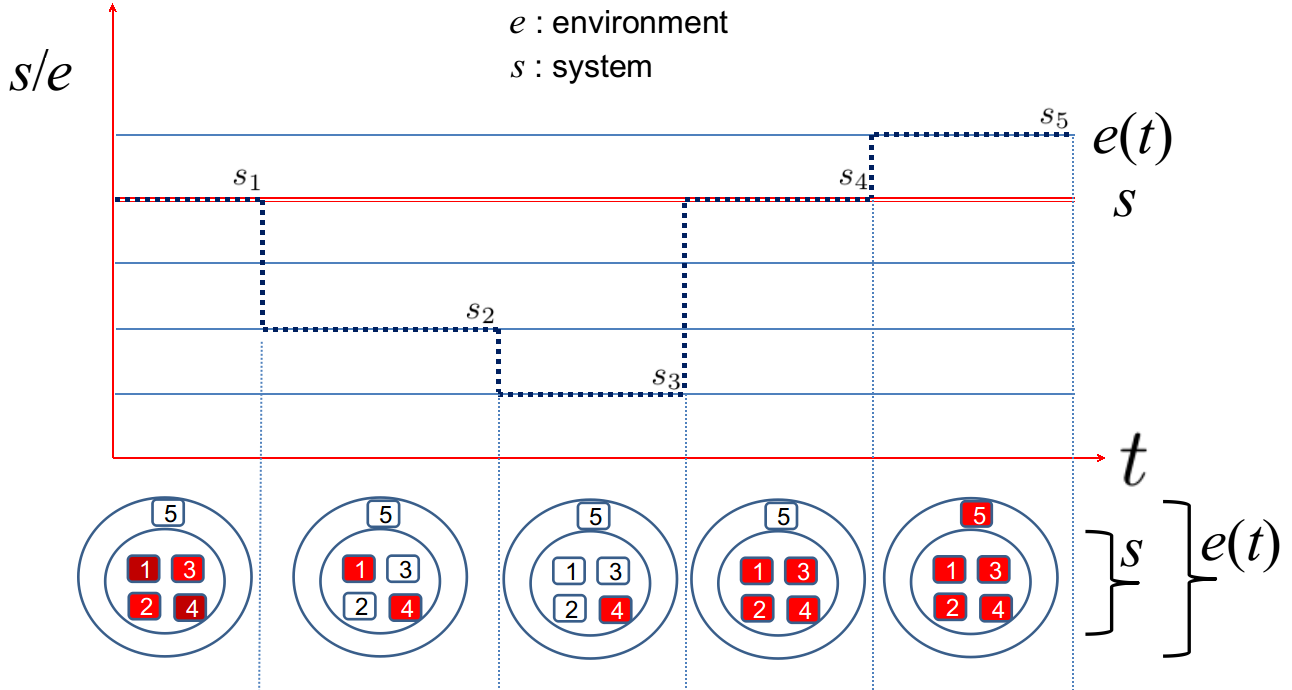}}
	\caption{Exemplification of supply and system-environment fit.}
	\label{f:fitset}
\end{figure*}

Figure~\ref{f:fitset} exemplifies a system-environment fit in the case of two behaviors $\beta^s$ and $\beta^e$
with $s\subsetneq e$. Environment $e$ affects five context figures identified by integers $1,\dots,5$ while $s$
affects context figures $1,\dots,4$.  The system behavior is assumed to be constant, thus
for instance
if $s$ is a perception organ then it constantly monitors the four context figures $1,\dots,4$.
On the contrary $\beta^e$ varies with time. Five time segments are exemplified ($s_1,\dots,s_5$) during
which the following context figures are affected:
\begin{description}
	\item[$s_1$]: Context figures $1,\dots,4$.
	\item[$s_2$]: Context figure $1$ and context figure $4$.
	\item[$s_3$]: Context figure $4$.
	\item[$s_4$]: Context figures $1,\dots,4$.
	\item[$s_5$]: Context figures $1,\dots,5$.
\end{description}
Context figures are represented as boxed integers, with an empty box meaning that the figure is not affected by the
behavior of the environment and a filled box
meaning the figure is affected. The behavior of the environment is constant within a time segment and changes
at the next one.
This is shown through the sets at the bottom of Fig.~\ref{f:fitset}:
for each segment $t_s\in \{s_1,\dots,s_5\}$ the superset is $e(t_s)$ while the subset is $s(s_t)$, namely $e(s_t)\cap s$.
The relative supply and the system-environment fit also change with the time segments.
During $s_1$ and $s_4$ there is perfect supply and best fit: the behavior exercised by the environment is evenly
matched by the features of the system. During $s_2$ and $s_3$ we have overshootings:
the systemic features are more than enough
to match the current environmental conditions. It is a case of
oversupply. Correspondingly, fit is rather low.
In $s_5$ the opposite situation takes place: the systemic features---for instance, pertaining to a perception organ---are
insufficient to become aware of all the changes produced by the environment. In particular here
changes associated with context figure 5 go undetected. This is a case of undersupply (that is to say,
undershooting), corresponding to a loss of identity:
the ``worst possible'' system-environment fit.

\subsection{Supply- and fit-aware behaviors}\label{s:or}

Whenever a partial order ``$\prec$'' exists
between a system's and its environment's behaviors
it is possible to consider system behaviors of the following forms:

\begin{enumerate}
	\item
	Either $b=\bhvrtwo{\hbox{pro}}{F}$ or $b=\bhvrtwo{\hbox{ant}}{F}$, with
$\sigma(b)=\hbox{false}$ and with $F$ including figures
that provide a measure of the risk of unresilience, expressed, e.g.,
through $\mathbf{supply}$ and $\mathbf{fit}$.\label{e:opt.pro}

Such behavior corresponds to condition~\ref{d:anti:sefOpen} in Def.~\ref{d:anti}, namely
one of the necessary conditions to computational antifragility.
When exercised by system organs for analysis, planning, and knowledge management,
this behavior
translates in the possibility to become aware of and speculate about the possible future resilience
requirements. If this is coupled with the possibility to revise one's system organs by
enabling or disabling, e.g., the ability to perceive certain context figures depending
on the extrapolated future environmental conditions, then a system could use this
behavior to improve proactively its own system-environment fit---possibly 
mutating its features and functions.

An exemplary system based on this feature is given by the already mentioned
adaptively redundant data structures of~\cite{DeBl08a} and
the adaptive $N$-version programming system introduced in~\cite{BDB11a,BDFB12}.
Those systems make use of so-called reflective variables~\cite{DB07a}
in order to perceive changes in a ``distance-to-failure'' function.
Such function basically measures the probability of failure of
a voting scheme at the core of the replication strategies adopted in the mentioned systems.
In other words, such function estimates the probability of
undersupply for voting-based software systems.

\item
	Behavior $b$ defined as in case~\ref{e:opt.pro},
	but with $\sigma(b)=\hbox{true}$.\label{e:opt.soc}

In this case the analysis, planning and knowledge organs are aware
of other systems in physical or logical proximity and may use this fact
to artificially augment or reduce their system features 
by establishing / disestablishing mutualistic relationships
with neighboring systems. An example of this strategy is sketched
in Sect.~\ref{s:sce}.
\end{enumerate}

Note how both behaviors~\ref{e:opt.pro} and~\ref{e:opt.soc} may evolve the system
beyond its current identity. In case~\ref{e:opt.soc} the behavior
augments the social ``scale'' of the system, which becomes
\emph{a part of a greater whole}---in other words, a resilient collective system~\cite{DF15a}.

As a final remark, we observe how the formulation of system-environment fit adopted in the
present article
may be augmented so as to include other factors---for instance, overheads and costs.

\section{Scenario}\label{s:sce}

In the present section the approach introduced in Sect.~\ref{s:app} is exemplified
through an ambient scenario.
As we did in~\cite{DF13b}, also our scenario here is inspired by the use of so-called
sentinel species~\cite{dS99+}, namely systems or animals able to compensate
for another species' insufficient perception.
We now introduce the main actors in our scenario.

\subsection{Coal Mine}
Our ambient is called ``Coal Mine''. Reason for this name is to highlight how
the behavior of this ambient may randomly change from a neutral state (NS) to a threatening state (TS),
as it occasionally occurs in ``real life'' coal mines when high concentrations of 
toxic substances---e.g., carbon monoxide and dioxide, or methane---manifest themselves.
(Toxic gases in high concentration are lethal to both animals and human beings.)

By making use of the terminology introduced in Sect.~\ref{s:bea:rwb} and Sect.~\ref{s:bea:pap}
we shall refer to the behaviors of Coal Mine as to $\bhvr{CM} = \bhvrtwo{ran}{T}$,
where $T$ represents a context set including a figure, let us call it $t$,
telling whether Coal Mine is in its neutral or threatening state.

Several systems may be deployed in Coal Mine. Let us call Miner one such system.

\subsection{Miner}
Miner is a system whose \emph{intrinsic\/} resilience (see Obs.~\ref{o:intr})
is very high: Miner's resilience
organs are capable of advanced behaviors, including perception of a wide range
of context figures; proactive analysis and planning; and a knowledge organ
able to persist lessons learned from experience.


In particular, let us refer to $\bhvr{M}$ as to the behavior of
the perception organ of the Miner
(that is to say, Miner.\M). Let us assume $\bhvr{M}$ to be a purposeful behavior
able to report changes in some set $F$ of context figures.
In other words, $\bhvr{M} = \bhvrtwo{pur}{F}$.

In what follows we assume $T\setminus\{t\} \subsetneq F$, and $t\notin F$. Those assumptions mean
that Miner.\M{} can become
aware of any type of changes in Coal Mine, with the exception of a NS-to-TS transition.
Miner is thus unable to perceive the threat and therefore it is unresilient with respect to Coal Mine.

\subsection{Canary}
Let us now suppose we have a second system called ``Canary''. Canary's organs are all
intrinsically inferior (cf. Obs.~\ref{o:intr})
with respect to Miner's, with the exception of its perception organ, Canary.\M.
Let us call $\bhvr{C}$ the behavior of Canary.\M. In what follows
we assume $\bhvr{C}$ to be equal to
$\bhvrtwo{pur}{G}$ for some set $G$ of context figures.
In addition, we assume that both $F \subsetneq G$ and $G \subsetneq F$ are false.
Miner.\M{} and Canary.\M{} are thus \emph{incommensurable}---none of the conditions
in Def.~\ref{d:partial.order}
apply: neither Miner.\M{} $\prec$ Canary.\M{} nor 
Canary.\M{} $\prec$ Miner.\M{} is true.

\subsection{Discussion}

The advanced features of the Analysis, Planning, and 
Knowledge organs of Miner allowed it to deduct two relevant facts:
\begin{enumerate}
\item $t\in G$. In other words, despite its comparably simpler nature, Canary can detect a NS-to-TS
transition in Coal Mine---what
in a ``physical'' coal mine would represent
a dangerous increase in the concentration of toxic gases.
\item Canary is more susceptible than Miner~\cite{Rabinowitz01012010}
to the persistence of the TS state in Coal Mine. In other words, when deployed in Coal Mine
in its threatening state, Canary is likely to experience general failures
(what we could refer to as ``total losses of the system identity'') much sooner than Miner.
\end{enumerate}

Miner thus realizes that,
by bringing along instances of the Canary system and by monitoring for their condition and
failures, it may artificially augment its perception organ. This technique is
known as biomonitoring~\cite{Tingey89}.
The new collective system Miner+Canary is now characterized by a perception organ
whose behavior is of type
$\bhvrtwo{pur}{F\cup G}$. Let us refer to Miner+Canary as to MC.

Now, as $t\in G$, it follows that $T\subsetneq F\cup G$. This matches one of the
conditions in Def.~\ref{d:partial.order}, thus
$\bhvr{CM} \prec \bhvr{MC.\M}$.
Behaviors are now commensurable, and it is possible to deduct that
Coal Mine now exhibits ``systemically inferior'' resilience
with respect to the monitoring organ of
MC.

It is now possible to define a strategy based on Miner and multiple
instances of Canary.

\subsection{Strategy}
First we estimate the supply of Miner.\M{} with respect to Coal Mine.
The estimation 
is based on a probabilistic assessment of the distance between the
two involved behaviors. Said assessment
is formulated by considering the amount of Canary replicas that have failed 
out of a predefined maximum equal to $|c|$:

\begin{tabbing}
\hspace*{3ex}\=\hspace*{2ex}\=\hspace*{2ex}\=\hspace*{2ex}\=\hspace*{2ex}\=\hspace*{35mm}\=\kill
\textbf{float EstimateSupply} (Coal Mine $cm$,
	Miner $m$,Canary $c$[\,])\\
\textbf{Begin}\\
\textsf{1}\>int $f$ = 0\\
\textsf{2}\>Query state of the Canary instances $c$ deployed in $cm$;\\
\textsf{3}\>For each failed Canary in $c$, increment $f$;\\
\textsf{4}\>return $|c|/2.0 -f$;\\
\textbf{End}\\
\end{tabbing}

Secondly, through the estimated supply we can derive an estimated fit as follows:
\begin{tabbing}
\hspace*{3ex}\=\hspace*{2ex}\=\hspace*{2ex}\=\hspace*{2ex}\=\hspace*{2ex}\=\hspace*{35mm}\=\kill
\textbf{float EstimateFit} (Coal Mine $cm$,
	Miner $m$,Canary $c$[\,])\\
\textbf{Begin}\\
\textsf{1}\>float supply = \textbf{EstimateSupply}($cm, m, c$[\,]);\\
\textsf{2}\>if supply $\ge 0$ then return $1/(1+\hbox{supply})$;\\
\textsf{3}\>else return FLOAT\_MIN;\\
\textbf{End}\\
\end{tabbing}

  \begin {figure*}
      \begin{center}
\includegraphics[width=0.7\textwidth]{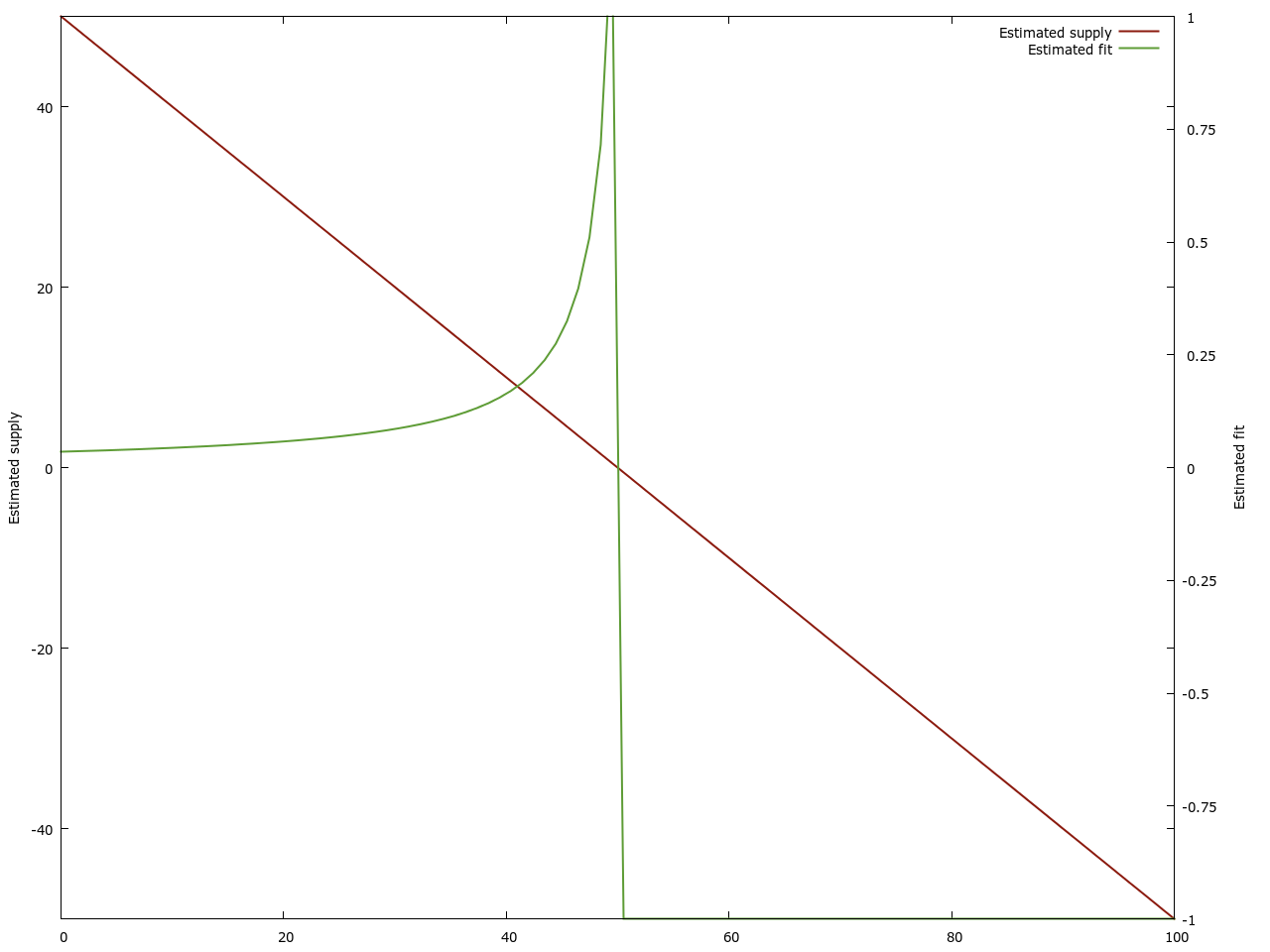}
\caption{Estimations of supply and fit when 100 instances of Canary are used. Abscissas represent
the number of failed Canary instances.}\label{f:supplyfit}
      \end{center}
    \end {figure*}

%

By executing a function like \textbf{EstimateFit}, Miner evolves into
a collective system by means of a \bhvr{\hbox{soc}}{} behavior.
Said behavior augments the system's \emph{social scale}, embedding the system into
a greater ``whole''.
In this case the established social relationship is parasitic rather than mutualistic,
as it enhances the resilience of one of the partners at the expenses of the other one.


\subsection{Final remarks}
The just described scenario is clearly an exemplary simplification. A full fledged example
would see a much more complex confrontation of behaviors of different organs and at
different system scales.
A possible conceptual framework for modeling such ``behavioral confrontations''
may be found in Evolutionary Game Theory~\cite{Sandholm2012} (EGT). In EGT
the system and its deployment environment could be modeled as opponents that
confront behavioral strategies of the classes discussed in this paper.
This interpretation matches particularly well cases where either or both
of the environment and the system explicitly aim at threatening their opponent,
as it is typical of certain security scenarios.

As a last remark, we observe how detecting incommensurability between
a system's behavior and its environment's provides the system with
awareness of the need to establish a social relationship---a new behavior
$b$ such that $\sigma(b)=$true---such as
a mutualistic or a parasitic relationship.



\section{Conclusions}\label{s:end}

This paper discusses resilience as the behavior resulting from the coupling of a system and its
environment. Depending on the interactions between these two ``ends'' and on the quality of the individual
behaviors that they may exercise, resilience may take the form of elasticity (change masking); entelechism
(change tolerance); or antifragility (adapting to and learning from change).
Following the lesson of Leibniz~\cite{DF14c}, resilience is decomposed in this paper into an intrinsic
and an extrinsic component---the former representing the static, ``systemic'' aspects
of a resilience design, the latter measuring the contingent match between that design
and the current environmental conditions.
It is conjectured that optimal resilience may be more easily attained through
behaviors that are not constrained by the hard requirement of
preserving the system identity.
Such ``antifragile behaviors'' are exemplified through a scenario in which a system
establishes a parasitic relationship with a second system in order to
artificially augment its perception capability.
Finally, we observe how several of the concepts discussed in this
paper match well with corresponding concepts in EGT. In
particular, the choice of which behavior to enact corresponds
with the choice of a strategy; resilience is the outcome
of an interplay---a ``game''; and the interplay between
the ``players'' (system and environment, as well as their
organs) translates in penalties and rewards.
Because of those similarities it is conjectured here that a possible
framework for the design of optimal resilience strategies may
be given through EGT, by modeling both system and
environment as two opponents choosing behavioral strategies
with the explicit purpose to ``win'' the adversary.
In this new model we shall distinguish between a system's behavior and a system's
\emph{manifested\/} behavior. The former is what we have
focused on in this paper and characterizes the ``systemic class'' of the system---what the system
is capable to do. The latter is the behavior the system \emph{decides\/} to manifest;
it is a ``move'' in a confrontation between two opponents.
Thus for instance an intelligent agent able to exercise advanced behaviors
may decide to behave (pseudo-)randomly so as to, e.g., confuse the opponent, or even to cause the opponent
choose a yielding point and then use this information to ``attack'' it and lead it
to an undershooting.
Future work will include proposing one such model
and assessing its benefits.

\subsection*{Acknowledgments}
I would like to acknowledge the valuable suggestions of the reviewers, which
allowed me to correct many inaccuracies and improve considerably the coherence
of this work. Last but far from the least I would like to thank my wife, father-in-law, and son, for
providing me with the environment, the time, and the tranquility
for me to concentrate on this article for several hectic weeks.

\bibliographystyle{spmpsci}

\end{document}